\documentclass[aps,prl,twocolumn,groupedaddress,amsmath,amssymb]{revtex4-1}
\usepackage{graphicx}  
\usepackage{dcolumn}   
\usepackage{bm}        
\usepackage{verbatim}   

\begin{document}

\title{The spatial statistics of turbulent dissipation rates}
\author{James Glimm}
\affiliation{
   Department of Applied Mathematics and Statistics,\\
   Stony Brook University,\\
   Stony Brook, NY 11794
   }
\author{Vinay Mahadeo}
\affiliation{
   Department of Applied Mathematics and Statistics,\\
   Stony Brook University,\\
   Stony Brook, NY 11794
   }
\date{\today}
\begin{abstract}
We study the spatial statistics of velocity gradient volatility 
(i,e., the energy dissipation rate) in turbulent flow.
We extend 
the Kolmogorov-Obukhov theory but also narrow its scope. The models are
log normal, with verification from finely resolved large eddy and
direct numerical simulations. They are parameterized by
a mean and a covariance operator. Addressing applications to large eddy simulations,
the mean and the covariance depend on the resolved scale solution. 
Removing this resolved scale dependence by a locally defined rescaling of
the turbulent statistics yields a universal theory for
the subgrid statistics, specifically for the mean and variance of 
turbulent fluctuations, i.e., the log of the energy dissipation 
rate $\epsilon$. 
The variance of the
velocity gradient statistics is found to be log normal, in accordance with
Kolmogorov 1962. Rescaling by the resolved scale mean and variance
removes the influence of the resolved scales on the subgrid
scales, justifying conceptually the universality we observe.
The universality is the basis of new power
scaling laws.
The new power laws, in turn, allow a simple parameterization of
the mean and covariance
of the subgrid rescaled dissipation rate, i.e., velocity gradient
volatility.
We treat the coarse grid
resolved space-time location as a random variable.
A restriction of the theory to a small range of unresolved but
stochastically modelled scales is proposed, with renormalization
group ideas offered to overcome this restriction.
 \end{abstract}

\maketitle

\section{Introduction}
The scaling law of Kolmogorov \cite{Kol41}, although simple, is among the
deepest contributions to the understanding of turbulent flow. Turbulent flow is
generally considered to be the major outstanding problem of classical physics.
We propose new parameterized spatial statistics and
scaling laws for turbulent fluctuations which extend
ideas of the Kolmogorov 1962 (K62) theory but also narrow its scope. Verification is
from comparison to Direct Numerical Simulation (DNS).
Because we do not assume 
homogeneity or isotropy
of the flow,
the theory is applicable to the unresolved scales of
Large Eddy Simulations (LES), and is
a stochastic subgrid scale (SGS) model. A major contribution of the theory
presented here is to capture SGS fluctuations, in addition to the means
computed from traditional SGS theories.

We assume single fluid, constant density, incompressible flow, governed by the
Navier-Stokes equation.
The scalar dissipation rate is defined as
\begin{equation}
\label{eq:dissipationDef}
\epsilon = \frac{\nu}{2}
\sum_{i,j}\left( \frac{\partial \mathbf{u}_i}{\partial \mathbf{x}_j} +\frac{\partial \mathbf{u}_j}{\partial \mathbf{x}_i} \right)^2 
= \frac{\nu}{2} \|S+S^*\|_2^2 \ ,
\end{equation}
with $\nu$ the dynamic viscosity, $\mathbf{u}$ the fluid velocity, and
\begin{equation}
\label{eq:strain_rate}
S = \frac{\partial \mathbf{u}_i}{\partial \mathbf{x}_j}
\end{equation}
the strain rate.

Properties of $\epsilon$ are fundamental to a number of deep
theories of turbulence, including
notions of turbulent intensity, turbulent intermittancy \cite{Sre04},
corrections to the Kolmogorov 5/3 exponent and scaling laws for
velocity moments. The analysis of $\epsilon$ has been applied to
turbulent diffusion, short distance asymptotics
for the velocity two point correlation function, and prediction of a fractal or
multifractal spatial distribution of regions of high turbulent intensity,
with the turbulent regions concentrated in a fractal set of dimension
$D < 3$ \cite{Cho94}. Scaling laws for wall bounded turbulence (a topic
not addressed here) are discussed in \cite{BarCho98}. There is an extensive
literature on the above topics, which we do not attempt to document
in this article.
See the survey article \cite{Jou97} for elaborations and further references.
Kolmogorov \cite{Kol62} and Obukhov \cite{Obu62}
conjectured that $\epsilon$ obeys a lognormal distribution at sufficiently 
high Reynolds numbers.

Our main results are new scaling law for the 
mean and covariance of $\chi = \ln{(\epsilon})$, 
which we designate as turbulent intensity.
The scaling laws originate
in a universality principle, of the mean $\mu$ and the
variance of $\chi$ for inertial
degrees of freedom. The connection between universality, which would
seem to require isotropic homogeneous turbulence, and subgrid scales
of LES, which can be very far from homogeneous or isotropic is one of
our central results. Depending on the local properties of the resolved
scales of turbulence, we introduce a rescaling of the turbulent statistics,
and it is only after this rescaling that the subgrid scales of turbulence
become universal. In fact, universality
does have limits, in terms of the
number of subgrid scales it can accommodate. In our numerical tests, we
consider 3 levels of mesh refinement, that is a change of length scales
by a factor of 8. For a large number of length scales, we have in mind
a renormalization group methodology, with successive integration over
smaller length scales, and a resetting of parameters using the
(statistical) knowledge of the already integrated scales, a point of view to
be elaborated in future work.

In Sec.~\ref{sec:random_field}, we specify the simulation study used
for verification of the theories proposed here. We define the
statistics and the ensemble used. We show that $\epsilon$
has log normal statistics (and $\chi$ has normal statistics).

The spatial statistics for
$\epsilon$ and $\chi$ are characterized by the mean 
$\mu$ and covariance $\Sigma^2$
of $\chi$. We compare a coarsely gridded 
LES to a finely gridded LES or DNS. The fine grid
extension of the coarse grid solution is of course unique and not stochastic,
but we add statistics in the dependence of this extension on the resolved
coarse grid cell from which it is derived. The statistics of the fine grid
extension for $\chi$ depends on the coarsely gridded resolved solution.

A detailed analysis of the $\chi$ statistics is
carried out in Fourier space,
in Sec.~\ref{sec:spatial-fourier}.
The analysis uses  the coarse grid resolved scales as
a random varible, to define the statistical properties of the subgrid
scales. 
Our premise is that the spatial aspects of this
dependence are largely captured by a resolved
scale variance $\Sigma_0^2$ and mean $\mu_0$ for $\chi$.
When the $\Sigma_0^2$ and $\mu_0$
dependency is removed from the
refined solution extension by rescaling, the resulting statistics for $\chi$,
i.e., $\Sigma^2/\Sigma^2_0$, is 
modeled as universal, with $\Sigma^2/\Sigma^2_0$ diagonal and
subject to simple modeling assumptions, including new power laws,
with only a few ${\cal{O}}(1)$ dimensionless parameters to govern 
its spatial statistics, all physically understandable.
The power laws result from a universality principal for the
rescaled statistics.
These approximations are confirmed by the fine grid comparison.

Sec.~\ref{sec:grad_vel}
extends this log normal
theory to individual components of the strain matrix $S$
and to the tensor dissipation rate.
A discussion of results and an outlook for future work is
given in Sec.~\ref{sec:conc}. 

\section{The spatial statistics}
\label{sec:random_field}

\subsection{The simulation study}
\label{sec:simulations}
All verification tests are based on a series of coarse and finely gridded
simulations.
The finely gridded  LES/DNS study has over two decades of resolution,
with the coarsely gridded LES fraction of the
solution varying with the simulation
parameters. The coarsely gridded LES includes portions of the inertial
range and has up to a decade of unresolved inertial range scales
to which our modeling applies.

Our fine and coarse grid simulations are
conducted with the code \cite{PouKolChe14}, using periodic boundary conditions
and forced turbulence in a domain size $L^3$, $L = 0.05m$, with the
viscosity (of air), $1.9e^{-5}m^2/s$. The mesh parameters for three
simulations are given in Table~\ref{table:mesh}. 
The resolved levels of turbulence in this numerical study
are not extreme.

Let $\Delta_c$ denote the coarse grid mesh size, and $\Delta_f$
the fine grid mesh size, so that $\Delta_c = 8\Delta_f$.
As we are concerned with velocity gradients, which require two mesh cells
as a minimum stencil size for their evaluation, we also introduce
what we call the resolved coarse and fine grid mesh scales, with
$\Delta_r = 2\Delta_c$ as the resolved coarse grid mesh and
\begin{equation}
\label{eq:mesh}
\Delta_r = 2\Delta_c = 8~ (\mathrm{Resolved}~ \Delta_f) = 16\Delta_f \ .
\end{equation}
Here the resolved fine grid level is $2\Delta_f$, needed to evaluate
the fine grid velocity gradients occurring in $\epsilon$ and $\chi$.

\begin{table}
\caption{
\label{table:mesh}
Mesh parameters for three fine grid and corresponding  coarse grid simulations
considered in this study.
}
\begin{center}
\begin{tabular}{|c|c|c|c|c|c|}
\hline
Fine grid & Fine and  coarse & Re  & Taylor Re & Kolmogorov  \\
simulation & grid meshes & & & scale \\
DNS & $256^3$ and $32^3$ &  561 & 40 & $1.4\times \Delta_f$ \\
LES & $256^3$ and $32^3$ & 1577 & 67 & $0.67\times \Delta_f$ \\
LES & $256^3$ and $32^3$ & 2539 & 85 & $0.04\times \Delta_f$ \\
\hline
\end{tabular}
\end{center}
\end{table}

\subsection{The statistical ensemble}
\label{sec:covariance}

As with turbulent flow in general, $\epsilon$ and $\chi$ 
can be regarded as random fields. In the LES context, 
$\epsilon$ and $\chi$ are known (deterministic) at the resolved scales,
while their subgrid scale fluctuations can be modeled using spatial statistics.
We write the fine grid resolved spatial mesh coordinate
$i = i_r,i_s$, where $i_r$ is the coarse grid
resolved mesh portion of $i$. As the
resolved coarse mesh spacing is $2\Delta_c$, $i_r$ spans all the coarse grid
LES cells other than the last or finest of them. $i_s$
is the remainder, occurring in the fine but not the coarse
LES grid resolution and also within the
finest level of the LES simulation. There are then $n_r = 16^3$ $i_r$ values and
$8^3$ $i_s$ values. Note that $2^3$ $\Delta_f$ mesh indices are missing,
as we discuss resolved, not simulation mesh indices.

Our plan is to treat $\chi$, regarded as a fiction of $i_s$, as random, 
depending on the random variable $i_r$.
The natural 
domain for its mean and covariance $\mu$ and $\Sigma^2$ is the
Hilbert space $\cal H$ of subgrid functions,
i.e., functions indexed by $i_s$ but restricted to a single resolved 
coarse grid cell
and having mean zero there. $\mu$ is a vector in $\cal{H}$ and
$\Sigma^2$ is an $i_s \times i_s$ positive definite matrix
acting on $\cal{H}$.

The statistics are collected as an average over all the possible $n_r$
values of $i_r$.

\subsection{The log normal property}
\label{sec:lognormal}

We present numerical evidence to support the hypothesis that
the fine grid extension of the
resolved $\chi$ defines a Gaussian random field, 
with the mean and covariance defined from
averages of distinct resolved coarse grid cells.

The time dependent spatial average of $\chi$ has been tested for normality,
see \cite{MonYag65,YeuPop89,PopChe90} and references cited there, based on experimental and 
DNS evidence.
%
Our verification method to establish the Gaussian property
is also based on DNS evidence.
A multivariate random variable is Gaussian if
its inner product with any fixed vector is univariate Gaussian. Here we 
consider, as representative choices, the
fixed vectors defined by single DNS cell indices $i_s$. 

We use QQ plots to assess univariate normality. In these, the trial distribution
is transformed first by a change of the $\chi$ independent
variables to have zero mean and unit variance.
Then we apply an inverse Gaussian change of trial distribution
dependent variables, with the 
result compared to a unit slope straight line.

\begin{figure}[ht!]
\centering
\includegraphics[scale=.13]{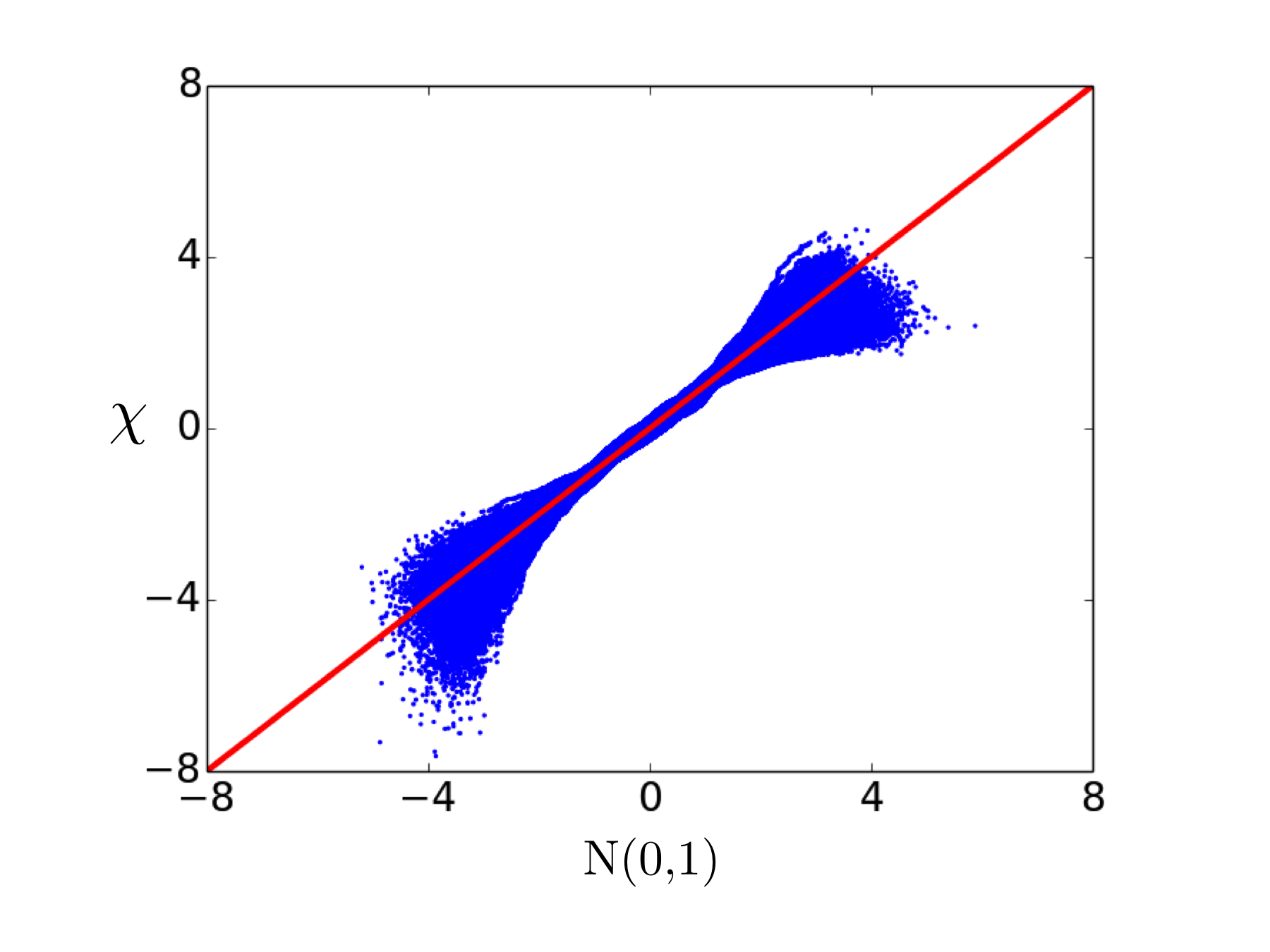}
\caption{QQ plot of the fine grid $\chi$ distribution in localized resolved 
cells compared to a normal distribution. Integral scale Re = 1577. 
Both axes are in units of
standard deviation. The agreement is good up to $\pm 2$ standard deviations,
which is the limit of accuracy imposed by the size of the data sample.
\label{fig:chiQQlocal}
}
\end{figure}

We create a single QQ plot by holding the fine grid mesh resolved cell indices
$i_s$ fixed ($16^3$ choices), and using the multiple
$i_r$ values from the $16^3$ resolved coarse grid cells to form the ensemble. 
In Fig.~\ref{fig:chiQQlocal} 
we superimpose all QQ plots for all choices of $i_s$
as typical tests of normality,
demonstrating agreement to a Gaussian law up to $\pm 2 \sigma$, 
a range of agreement consistent with the 4\% accuracy expected
from the ensemble size.

\subsection{The mean and covariance}
\label{sec:mean-variance}

The coarse grid resolved level variance of $\chi$ is  basic to our rescaling
strategy. 
At the level of each coarse grid resolved grid cell, the variance
$\Sigma_0^2$ of $\chi(i_r)$ is defined as

\begin{equation}
\Sigma_0^2 = (\chi(i_r) - \mu_0)^2
\end{equation}
with the definitions 
$\mu_0 = \overline{\chi(i_r)} = n_r^{-1}\sum_{i_r} \chi(i_r)$.

Consider the fine grid values $\chi(i_r,i_s)$ for the resolved coarse
grid $\chi(i_r)$.
As a normal random variable, the statistics of $\chi$ is characterized
by its mean and its covariance. To obtain universal statistics,
we rescale $\chi$ by subtracting the resolved coarse grid mean
and dividing by the standard deviation
defining $\widehat{\chi}(i_r,i_s) = (\chi(i_r,i_s) - \mu(i_s))/\Sigma_0(i_r)$.
The mean $\mu$ of $\chi$
is 

\begin{equation}
\label{eq:mu-hat}
\mu(i_s) = n_r^{-1}\sum_{i_r}\chi({i_r,i_s}) \ .
\end{equation}

The universal, rescaled, statistics are the statistics of
$\widehat{\chi}$. 
We follow a similar definition for the covariance, $\widehat{\Sigma}^2$ of $\widehat{\chi}$,

\begin{equation}
\label{eqn:covariance}
 \widehat{\Sigma}^2(i_s,j_s)
= \overline{ \widehat{\chi}(i_r,i_s)\widehat{\chi}(i_r,j_s)}  \ .
\end{equation}

\noindent where the overbar denotes averaging over all resolved coarse grid cells, namely

\begin{equation}
\overline{\left( \cdot \right)} = n_r^{-1}\sum_{i_r}\left( \cdot \right) \ .
\end{equation}

Furthermore, we will normalize $\widehat{\Sigma}^2$ by $\widehat{\Sigma}^2(i_r,i_r) = \widehat{\Sigma}^2(i_r) = \overline{\widehat{\Sigma}_0^2}$, the resolved scale variance, reducing the Reynolds number dependence of $\widehat{\Sigma}$.

\section{Spatial statistics in Fourier space}
\label{sec:spatial-fourier}

We propose a Fourier space analysis for $\mu$ and $\widehat{\Sigma}^2$.
As these variables are independent of the resolved variables,
they are, in this approximation,  periodic within each resolved
cell. Thus their Fourier transforms depend on the
Fourier modes within a single resolved cell.

In preparing figures in this section, we present functions of $\mathbf{k}$,
unless otherwise noted,
after binning all $\mathbf{k}$ values into bins 
labeled by the scalar wave number
$k$ with 200 bins whose size is linear in $\ln k$.

\subsection{$\mu$ in Fourier space}
\label{sec:mu}

\begin{figure}
\includegraphics[scale=.41]{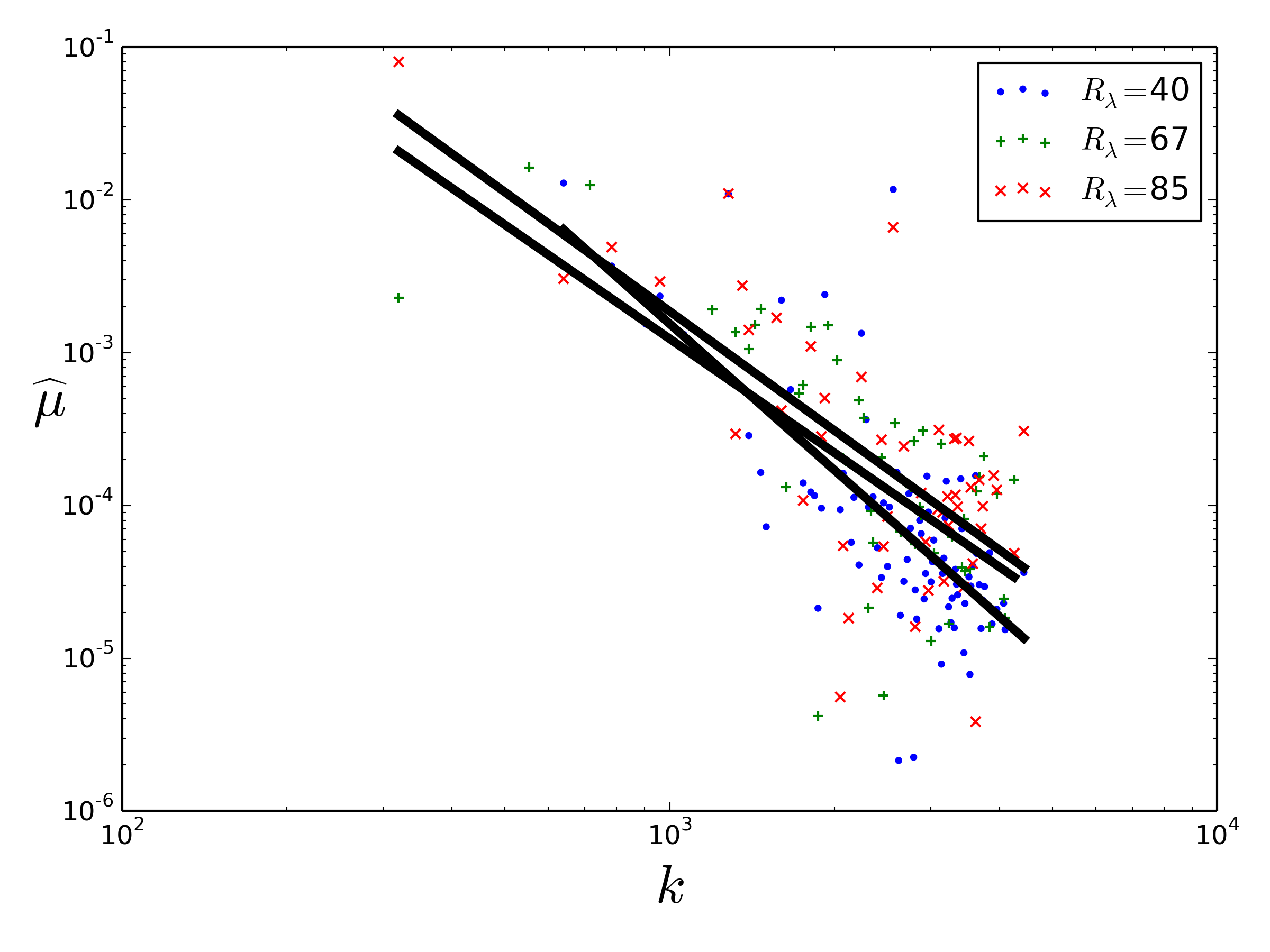}
\caption{
\label{fig:mu}
Plot of $\widehat{\mu}(k)$ vs. $k$ (unbinned) for three $R$ values.
The few values of $\widehat{\mu}$ less than zero are omitted.
}
\end{figure}

In Fig.~\ref{fig:mu}, we plot $\mu(k)$ vs. $k$ in
$\ln$ $\ln$ scaled variables.
Observe the approximate scaling law $\mu(k) \sim k^{-d}$, with corrections coming from
the dissipation range of scales. Included in the plot is a model
curve of the form
\begin{equation}
\label{eq:mean-model}
\mu(k) \sim k^{-d}
\end{equation}
The coefficients in this model fit are tabuleted in
Table~\ref{table:mean-model}.

\begin{table}
\caption{
\label{table:mean-model}
Coefficients for the mean model scaling law, for three values of $Re$.
}
\begin{center}
\begin{tabular}{|l|c|c|c|}
\hline
$Re$ & 561 & 1577 & 2539 \\
\hline
$d$ & $3.1$ & $2.5$ & $2.6$ \\
\hline
\end{tabular}
\end{center}
\end{table}

\begin{table}
\caption{
\label{table:var-model}
Coefficients for the variance model scaling law, for three values of $Re$.
}
\begin{center}
\begin{tabular}{|l|c|c|c|}
\hline
$Re$ & 561 & 1577 & 2539 \\
\hline
$C_2$ & $5.3$ & $3.9$ & $4.7$ \\
$C_4$ & $2.6$ & $5.3$ & $7.3$ \\
\hline
\end{tabular}
\end{center}
\end{table}

\begin{table}
\caption{
\label{table:matrix-model}
Representative coefficients of the scaling laws for the dissipation matrices.
}
\begin{center}
\begin{tabular}{c|ccc|ccc|}
		    \multicolumn{1}{c}{} & \multicolumn{6}{c}{$S_{ij}^2$} \\ \cline{2-7}
		    & \multicolumn{3}{c|}{$i=j$} & \multicolumn{3}{c|}{$i\neq j$}\\ \cline{2-7}
		    $Re$ & 561 & 1577 & 2539 & 561 & 1577 & 2539 \\
		    $C_2$ & $4.9$ & $3.9$ & $3.7$ & $5.0$ & $4.5$ & $3.1$ \\
		    $C_4$ & $1.2$ & $2.0$ & $2.3$ & $1.4$ & $2.3$ & $2.6$ \\ \cline{2-7}
			\multicolumn{7}{c}{}\\
		    \multicolumn{1}{c}{} & \multicolumn{6}{c}{$S^*S$} \\ \cline{2-7}
		    $C_2$ & $2.6$ & $4.1$ & $4.2$ & $4.6$ & $3.7$ & $4.0$ \\
		    $C_4$ & $1.4$ & $2.6$ & $3.2$ & $1.4$ & $2.9$ & $3.6$ \\ \cline{2-7}
			\multicolumn{7}{c}{}\\
		    \multicolumn{1}{c}{} & \multicolumn{6}{c}{$SS^*$} \\ \cline{2-7}
		    $C_2$ & $4.8$ & $4.2$ & $3.2$ & $4.4$ & $4.6$ & $3.9$ \\
		    $C_4$ & $1.4$ & $2.6$ & $3.6$ & $1.4$ & $2.9$ & $3.6$ \\ \cline{2-7}

\end{tabular}
\end{center}
\end{table}

\subsection{$\Sigma^2$ in Fourier space}
\label{sec:sigma}

\begin{figure}
\centering
    \includegraphics[scale=.41]{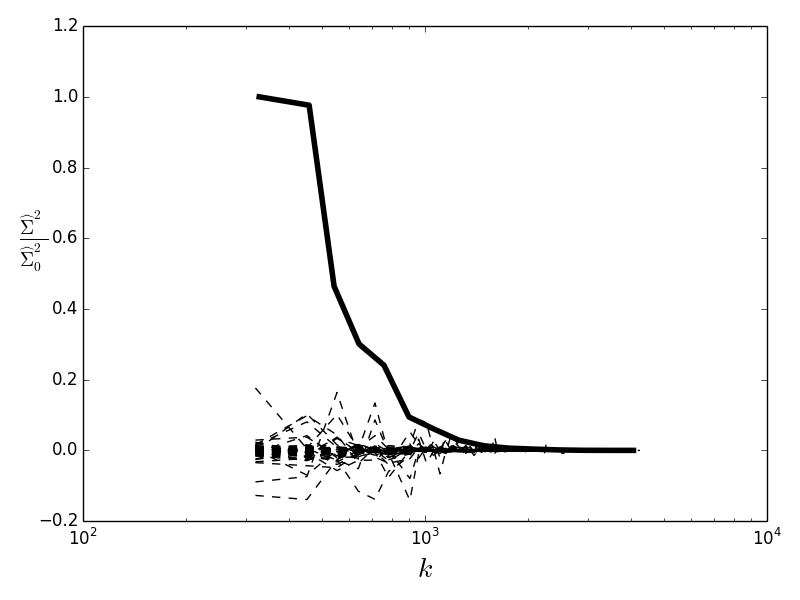}
\caption{
Plot of diagonal and off diagonal elements of the
covariance $\widehat{\Sigma}^2(k,l)$
at $Re=2539$ 
as a function of frequency $(k+l)/2$.
We observe a near vanishing of the off diagonal elements 
vs. the diagonal elements.  The same behavior is observed 
for the other two Reynolds numbers, but omitted from this plot.
\label{fig:covdiag}
}
\end{figure}

In Fig.~\ref{fig:covdiag}, we plot the diagonal and off diagonal elements of
the covariance matrix 
$\widehat{\Sigma}^2(\mathbf{k},\mathbf{l})$ for Reynolds number 2549.
The smooth curve in the upper part of the figure
represent the plot of the diagonal elements as a 
function of the frequency $k$, while the noisy data near
the bottom of the plot are the collection of all off diagonal elements,
plotted vs. $(k+l)/2$. The other Reynolds numbers give similar plots.
We conclude that the covariance is, within a good
approximation, diagonal.

Our modeling for $\widehat{\Sigma}^2$ (and for related operators describing the
strain matrix $S$ and its individual matrix entries in 
Sec.~\ref{sec:grad_vel}) is based on a simple approximation, which appears
to be consistent with the DNS data as examined here. We note that the
two random variables contributing to the covariance $\Sigma^2$ are
orthogonal for $\mathbf{k_s} \neq \mathbf{l}_s$ and so the 
$\mathbf{k}_s \neq \mathbf{l}_s$ terms vanish. In other words, the Fourier
coefficient of $\Sigma^2(i_s,j_s)$ is $\Sigma^2(\mathbf{k_s},\mathbf{k}_s)$,
which we write as $\Sigma^2(\mathbf{k}_s)$.  In our DNS case,
$\widehat{\Sigma}^2$ is a $16^3 \times 16^3$ matrix before binning.

The Fourier coefficients for $\widehat{\chi}$ for distinct angular variables
$\mathbf{k}/\|\mathbf{k}\|$ are statistically independent. For distinct
values of the scalar
Fourier amplitude $k = \|\mathbf{k}\|$,
we also assume statistical independence. These assumptions lead to a simple model for 
$\widehat{\Sigma}^2$, 
tractable for practical use.

We regard the wave number $k$ as an observable on 
the random field $\widehat{\chi}$, with the variance 
$\widehat{\Sigma}^2(k)$ for
this observation. We derive $k$ dependent scaling laws.
Approximately, the subspace of the Hilbert space in which
$\widehat{\Sigma}^2 (k)$ is defined has dimension proportional to $4 \pi k^2$,
namely the area of
a sphere in momentum space of radius $k$. Assuming equipartion of 
the variance of $\chi$ within this subspace, 
the inverse covariance is a multiple of the
identity there. We extend the variance equipartition hypothesis,
meaning a uniform level of variance,
to distinct $k$ values. 
The subspace variance is then proportional to $k^{-2}$.

From a more probabilistic point of view, we regard the observation as the
result of asking ${\cal{O}}(k^2)$ questions, postulated to be independent:
what is the variance of
$\chi$ associated with each $\mathbf k$ lying on the sphere of radius $k$?
Combining the variance of ${\cal O}({k^2})$ 
independent quantities gives the same answer. 

Volume scaling laws apply to the volume dependent coefficients in
the scaling law for $\widehat{\Sigma}^2$, and use a similar reasoning.
Thus, again assuming
equipartition, the mesh cell averages over a single coarse grid resoled
cell scale with $\Delta_r^3$.
In summary, we propose scaling laws 
for $\widehat{\Sigma}^2(k)dk$ in terms of
the scalar wave number $k$ and spatial 
volume, with the dispersion relation
\begin{equation}
\label{eq:dispersion}
\frac{\widehat{\Sigma}^2(k)}{\overline{\widehat{\Sigma}_0^2}} = 
\left( \Delta_r^3 4 \pi C_2^2 k^2 \exp \left(C_4^2 \eta^2 k^2\right) \right)^{-1} \,
\end{equation}
with dimensionless constants $C_2$, $C_4$ to be determined. 
The factors $\Delta_r^3$ and $4\pi k^2dk$ are volumes for physical space
and Fourier space. The $C_4$ term,
not previously discussed, is postulated to capture the leading order
viscous corrections to the otherwise purely inertial range theory.

\begin{figure}[ht!]
    \centering
    \includegraphics[scale=.4]{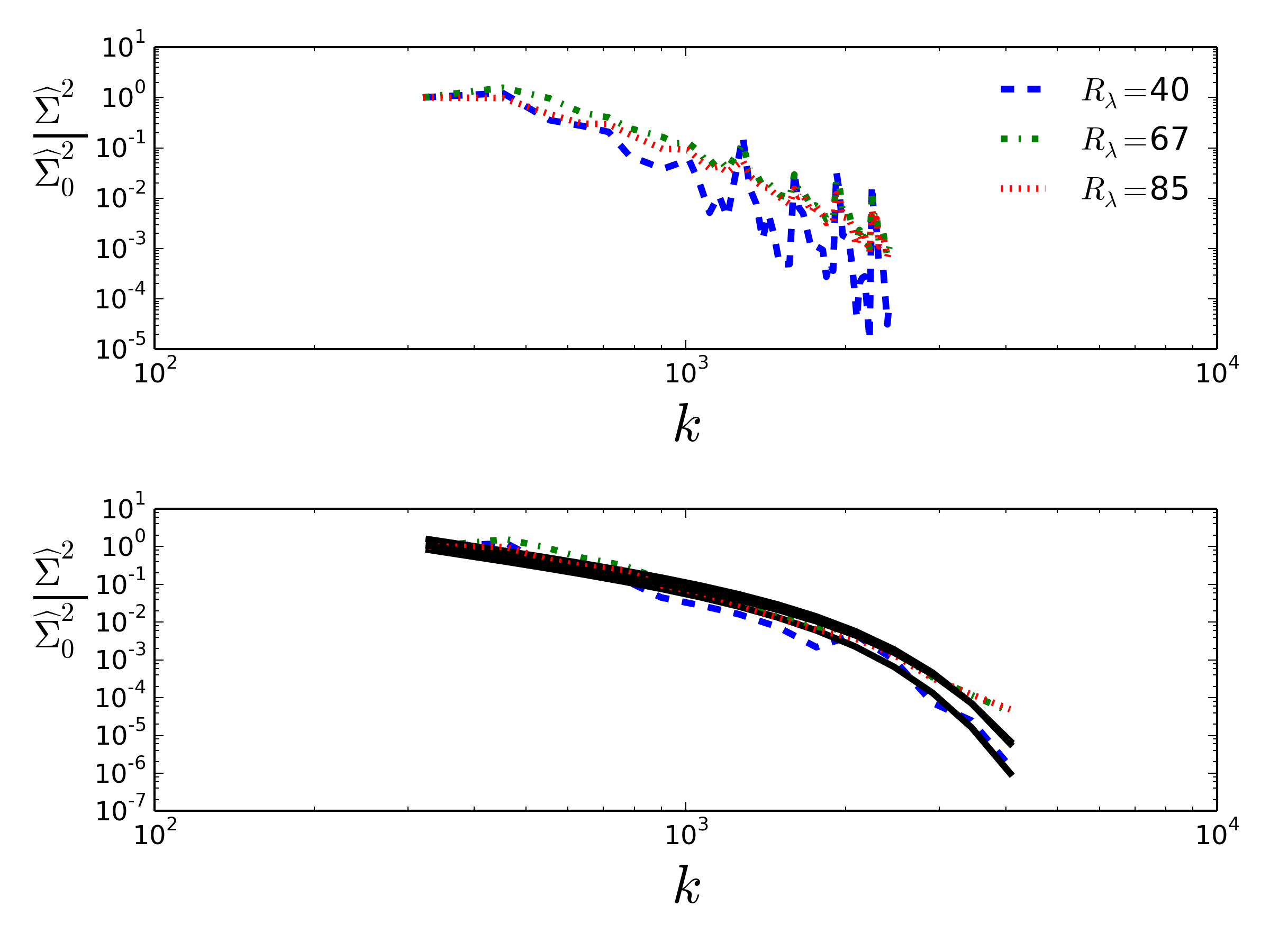}
    \caption{$\ln - \ln$ plot of the diagonal elements of
    $\widehat{\Sigma}^2(k)$ vs. $k$ at different Taylor scaled
    Reynolds numbers as a function of frequency. Top: unbinned curves.
	Bottom: binned curves. Solid black lines represent
    the fitting of the form (\ref{eq:dispersion}). 
    }
    \label{fig:chiVarRe}
\end{figure}

In Fig.~\ref{fig:chiVarRe}, we plot
$\widehat{\Sigma}^2(k)$ vs. $k$
on $ln$ $ln$ scales with the
model curves (\ref{eq:dispersion}) superimposed, based on optimal choices
of the coefficients $C_i$. We first fit $C_2$ in the inertial range
(small $k$ values), and then fit $C_4$ over a 1/2 decade at the end of the
interial range. We observe that most of the Reynolds number
dependence of the $C_i$ 
has been removed by the scaling introduced here, see 
Table~\ref{table:mean-model}. 

\section{A log normal model for velocity gradients}
\label{sec:grad_vel}

\begin{figure}[ht!]
    \centering
    \includegraphics[scale=.30]{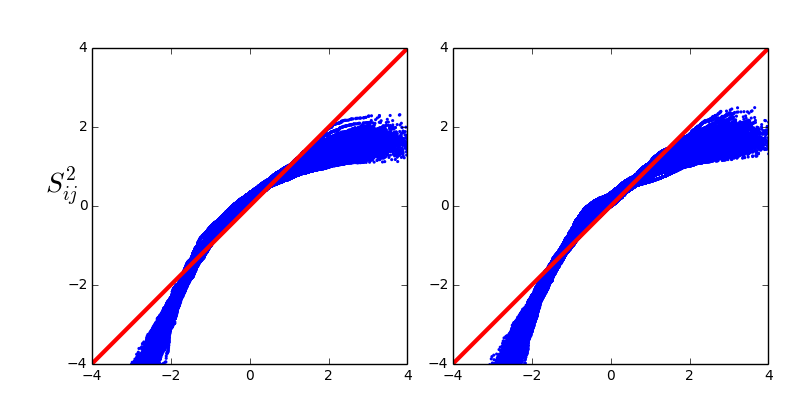}
    \caption{Two representative QQ plots for the variables $\ln \epsilon_{1,2}$ 
    corresponding to a diagonal (left) and off-diagonal (right) element the matrix $S$.
    As in Fig.~\ref{fig:chiQQlocal},
    the agreement with  a log normal distribution
    is satisfactory up to $\pm2\sigma$.}
    \label{fig:Sqq}
\end{figure}

We repeat the analysis of Secs.~\ref{sec:random_field} 
-- \ref{sec:sigma} for the strain matrix restricted to subspaces. Let $E_1$
and $E_2$ be self adjoint projection operators on $R^3$, and consider the
reduced strain matrix $E_1 SE_2 = S_{1,2}$.
In more detail, we consider 
\begin{equation}
\label{eq:dissipation-i}
\epsilon_{1,2} =
( \nu / 2 ) \| E_1\frac{\partial\mathbf{u}_i}{\partial x_j} E_2 \|_2^2 
\end{equation}
and the associated 
$\chi_{1,2} = \ln (\epsilon_{1,2} )$.
Thus $S_{1,2}$ is the strain associated with fluid velocities in directions
spaned by $E_1$ and fluctuations occurring with spatial variation in 
directions spanned by $E_2$.

\begin{figure}[ht!]
	\centering
	\includegraphics[scale=.45]{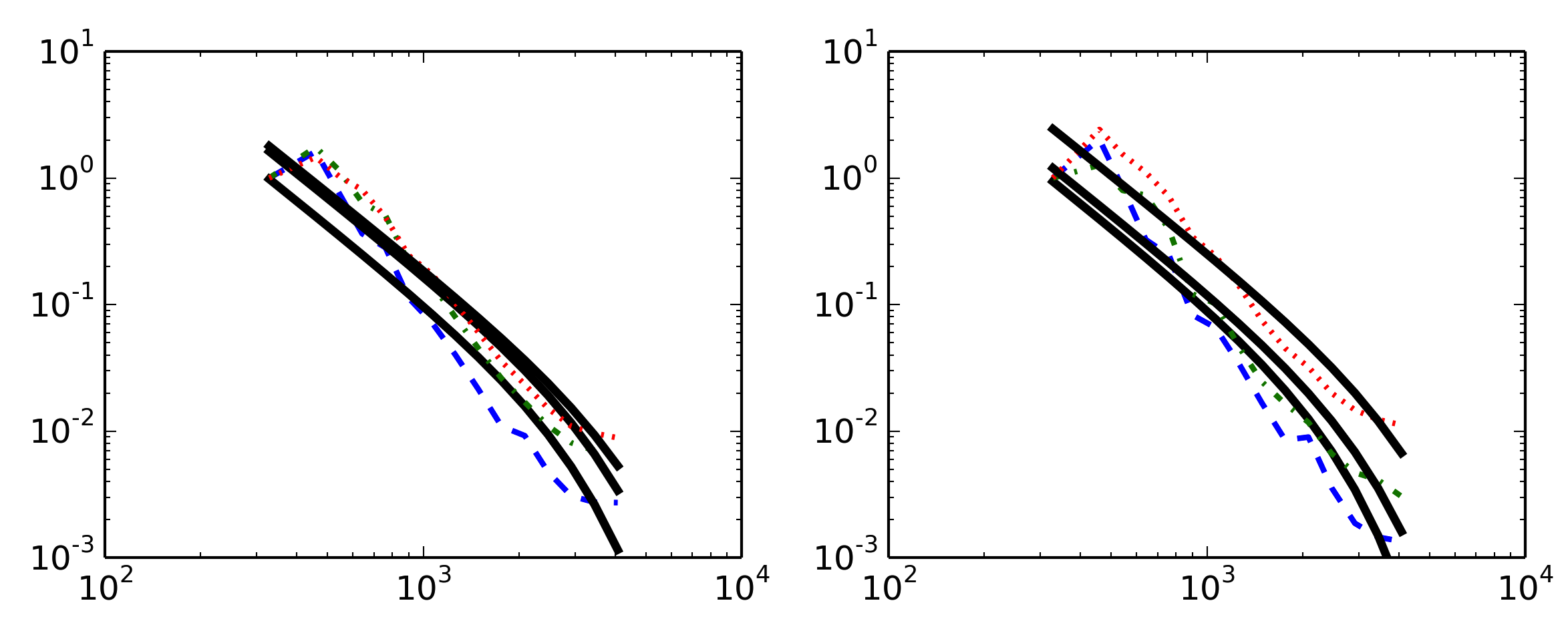}
	\caption{The diagonal elements of 
	$\Sigma^2_{1,2}(k)/\overline{\Sigma^2_{0,1,2}}$ plotted
	vs. $k$ in $\ln$-$\ln$ scales for two representative 
    matrix elements of $S_{1,2}$. 
	An $i=j$ case is presented on the left and
	an $i\neq j$ case presented on the right.
	Solid lines are the fitted curves defined in (\ref{eq:dispersion-i}).}
\end{figure}

With $\widehat{\mu}_{1,2}$ the rescaled resolved grid level mean for 
the rescaled $\widehat{\chi}_{1,2}$, 
$\widehat{\Sigma}^2_{1,2}$ is the rescaled covariance 
\begin{equation}
\label{eq:covariance-i}
\begin{aligned}
& \widehat{\Sigma}_{1,2}^2(i_s,j_s) =\\
& n_r^{-1} \sum_{i_r} (\widehat{\chi}_{1,2}(i_r,i_s) -\widehat{\mu}_{1,2}(i_r)) 
(\widehat{\chi}_{1,2}(i_r,j_s) - \widehat{\mu}_{0,1,2}(i_r)) \ ,
\end{aligned}
\end{equation}
With these changes, we repeat the analysis of 
Figs.~\ref{fig:chiQQlocal}--\ref{fig:chiVarRe}
with the modified dispersion relation
\begin{equation}
\label{eq:dispersion-i}
\begin{aligned}
&\widehat{\Sigma}^2_{1,2}(k)= \\
& \left (\mathrm{dim}~E_1~\mathrm{dim}~E_2\Delta_r^3 4 \pi C_{2,1,2}^2 k^2 \exp \left( C_{4,1,2}^2 \eta^2 k^2 \right) \right)^{-1} \ ,
\end{aligned}
\end{equation}
and modified coefficients $C_{2,1,2}$ and $C_{4,1,2}$. 
We assume that $E_1$ and $E_2$ are both one dimensional.

\begin{figure}
\includegraphics[scale=.41]{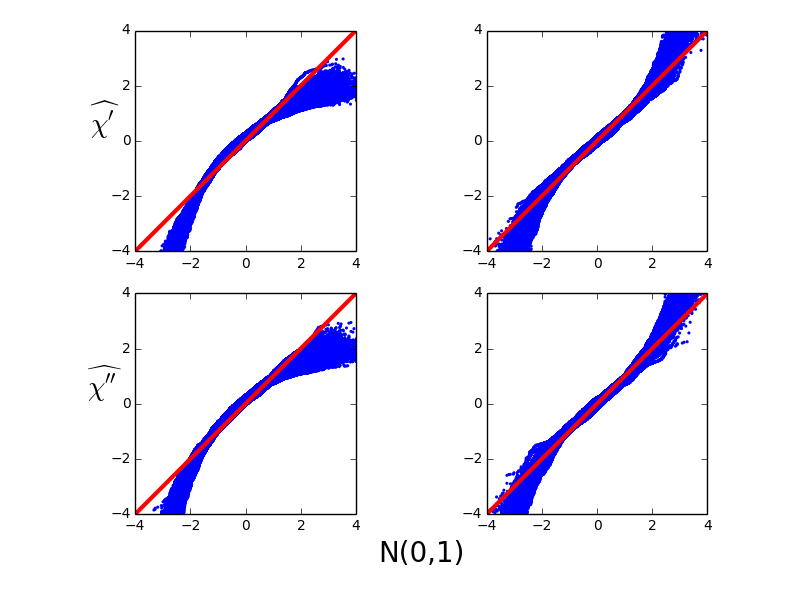}
\caption{
\label{fig:matrix_lognormal}
The normal statistics property for representative tensor components, diagonal 
and off diagonal, of the $\chi'$ and $\chi''$ tensors.
}
\end{figure}

\begin{figure}
\includegraphics[scale=.4]{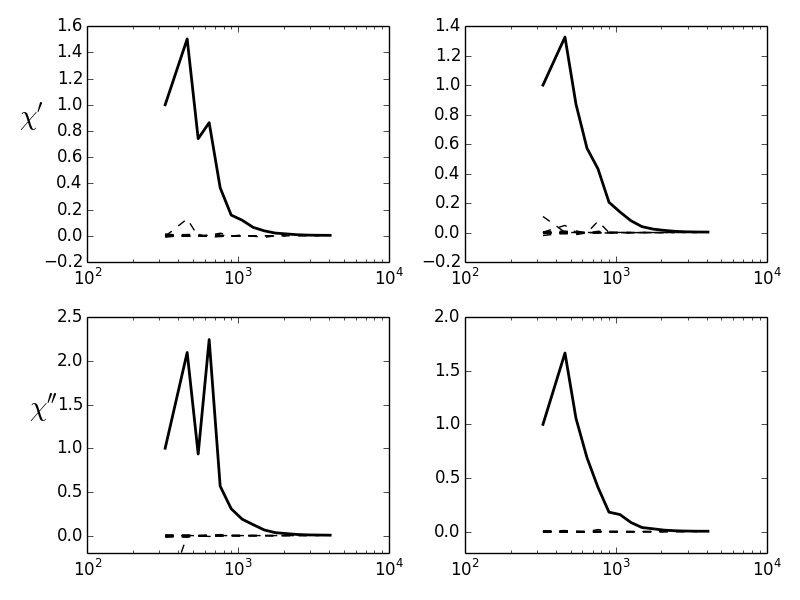}
\caption{
\label{fig:matrix_diagonal}
The Fourier diagonal property for representative tensor components of the
$\chi'$ and $\chi''$ statistics. 
The two plots on the left are the diagonal components and the two on the right are off-diagonal components.
}
\end{figure}

\begin{figure}
\includegraphics[scale=.4]{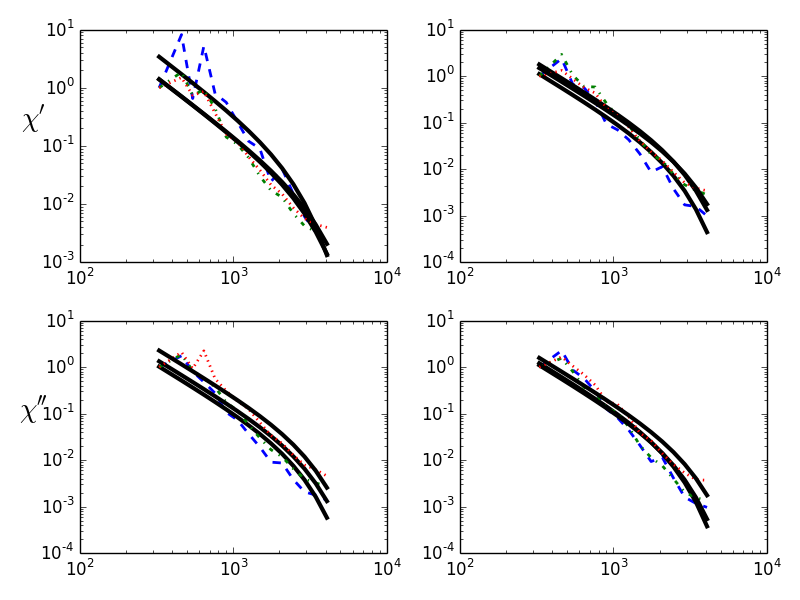}
\caption{
\label{fig:matrix_dispersion}
The dispersion relation for representative tensor components of the $\chi'$ and
$\chi''$ statistics.
The two plots on the left are the diagonal components and the two on the right are off-diagonal components.
}
\end{figure}

We also show that the tensor dissipation operators
\begin{equation}
\label{eq:epsilon'}
\epsilon' = (\nu/2) S^*S \ , \quad \quad \epsilon'' = (\nu/2)SS^*
\end{equation}
satisfy log normal statistics. Here the exponential and the log
in the definition of log normal 
($\chi' = \log(\epsilon')$ and
$\chi'' = \log(\epsilon'')$) 
is considered in the sense of matrix operations.
See Figs.~\ref{fig:matrix_lognormal},
\ref{fig:matrix_diagonal} and \ref{fig:matrix_dispersion}.

We note that the matrix QQ plots, and to some extent, all the QQ plots, raise
the possibility of nonGaussian statistics as a correction to the normal and
log normal statistics assumed here.
Especially in Fig.~\ref{fig:matrix_lognormal},
systematic corrections to the Gaussian property show smaller positive
excursions and stronger negative excursions for $\chi$
than Gaussian for the diagonal
matrix elements. Both the negative and positive differential
excursions reflect a smaller $\epsilon$ and
a possible bias toward a laminar rather than turbulent flow. In contrast, the
off diagonal QQ plots suggest stronger $\chi$ excursions than Gaussian
for both positive and negative excursions,
a phenomena known as heavy tails in the statistical data analysis literature.
These are the matrix elements which occur in vortex stretching and folding,
commonly believed to be an important element of fully developed turbulence.

\section{Discussion}
\label{sec:conc}

Verification by computational studies will typically encounter only
a limited range of LES unresolved scales, compared to a finely gridded
simulation. But the previous analysis, when applied over many
scales, has a conceptual flaw. If there is no resetting of the
$\Sigma_0^2$ parameters within the statistics model of a single
resolved scale, then we could apply this theory in the absence of
the coarsely gridded LES to the entire computation. Doing this would
remove the rescaling already seen to be important. For this reason, we
restrict the model to a small range of unresolved scales. In 
a future publication 
we reformulate the model in a renormalization group
manner in the case of multiple scales.

In an LES context, we analyze the unresolved scales statistically, with 
an emphasis on the strain rate $\epsilon$ and its logarithm 
$\ln \epsilon = \chi$. Resolved scale adjustments to the mean
and variance for $\chi$ lead to a theory for the resulting covariance $\Sigma^2$.
We find that $\Sigma^2$ is approximately diagonal in Fourier space.
An equipartition hypothesis for the variance of $\chi$ 
leads to a new power scaling law and 
a simple parameterization for $\Sigma^2$, verified by comparison to DNS
over about a decade of LES unresolved inertial scales. For reasons discussed,
we restrict this theory to a single or limited range within the inertial
range of scales.
This pure dissipation rate model is extended to the individual velocity
gradient components
$\epsilon_{1,2} = (\nu / 2) (\partial \mathrm{u}_i / \partial x_j)^2$
and to the tensor dissipation rates $S^*S$ and $SS^*$.

Open questions, to be addressed in future work, concern temporal statistics of
turbulent intensity, statistical models for the velocity gradients and 
strongly nonhomogeneous flows, such as boundary layers.

\bibliographystyle{plain}
\bibliography{refs}

\end{document}